# Metrics and Mechanisms:
# Measuring the Unmeasurable in the Science of Science

**Highlights**

We introduce, compare, and discuss recently proposed contextualized and theory-derived science metrics.

We demonstrate how emerging metrics complement existing ones in accessing the performance of scientists and teams, expose hidden structures and mechanisms of science, and provide additional theoretical insights illuminating the foundations of the science of science.

We group metrics and theories by three fundamental properties of science: hot and cold science, soft and hard science, fast and slow science to remind us that understanding and modeling the mechanisms of science condition effective development and application of metrics.

# Metrics and Mechanisms:
# Measuring the Unmeasurable in the Science of Science

Lingfei Wu[1*]   Aniket Kittur[2]   Hyejin Youn[3,4]   Staša Milojević[5]   Erin Leahey[6]   Stephen M. Fiore[7]   Yong Yeol Ahn[5*]

**Abstract**

What science does, what science could do, and how to make science work? If we want to know the answers to these questions, we need to be able to uncover the mechanisms of science, going beyond metrics that are easily collectible and quantifiable. In this perspective piece, we link *metrics to mechanisms* by demonstrating how emerging metrics of science not only offer complementaries to existing ones, but also shed light on the hidden structure and mechanisms of science. Based on fundamental properties of science, we classify existing theories and findings into: *hot and cold* science referring to attention shift between scientific fields, *fast and slow* science reflecting productivity of scientists and teams, *soft and hard* science revealing reproducibility of scientific research. We suggest that interest about mechanisms of science since Derek J. de Solla Price, Robert K. Merton, Eugene Garfield, and many others complement the zeitgeist in pursuing new, complex metrics without understanding the underlying processes. We propose that understanding and modeling the mechanisms of science condition effective development and application of metrics.

[1] School of Computing and Information, The University of Pittsburgh, 135 N Bellefield Ave, Pittsburgh, PA 15213
[2] Human-Computer Interaction Institute, Carnegie Mellon University, 5000 Forbes Ave Pittsburgh, PA 15213
[3] Kellogg School of Management, Northwestern University, 2211 Campus Dr, Evanston, IL 60208
[4] Santa Fe Institute, 1399 Hyde Park Rd, Santa Fe, NM 87501
[5] Luddy School of Informatics, Computing and Engineering, Indiana University Bloomington, 919 E 10th St. Bloomington, IN 47408
[6] School of Sociology, University of Arizona, 1145 E. South Campus Drive, Tucson, AZ 85721
[7] Department of Philosophy and Institute for Simulation and Training, University of Central Florida, 3100 Technology Parkway, Orlando, FL 32826

\* Corresponding authors: Lingfei Wu (liw105@pitt.edu) and Yong Yeol Ahn (yyahn@iu.edu)

# 1. Introduction

**Linking Metrics and Mechanisms**

What science does, what science could do, and how to make science work? If we want to know the answers to these questions, we need to be able to uncover the underlying mechanisms of science and not just focus on creating the metrics that are easy to collect and quantify. Metrics can sometimes deceive us. For example, the decline of citations to a scholarly work may not be because it is no longer relevant or important, but because the once-novel idea is now common knowledge and does not need any explicit reference—we don't cite Galileo Galilei to argue that the earth orbits around the Sun instead of the reverse (Merton 1968; McCain 2014). The observation that the most cited work of a scholar is more likely to appear earlier rather than later in their career leads to the enduring myth of young, creative scientists. Sinatra et al. (2016) have shown that this myth could be an illusion created by the uneven distribution of research resources over a career—young scholars have more time to produce, and old scholars have the social and intellectual capital to disseminate. After controlling for productivity, the most influential work may happen at any stage of a scientist's career (Sinatra et al. 2016). The exponential increase of scholarly literature has been used to celebrate the growth of total knowledge. Still, we often forget that this exponential pace depends on our choice of the unit of analysis to measure the knowledge stock. If we look at new concepts rather than new articles, the cognitive content of science expands slowly, following a linear form (Milojević 2015). These are just a few cases among many more in which metrics can be misleading or come to be misused when there is a lack of understanding of the mechanisms that produce them (Chen 2016).

In this perspective piece, we link *metrics to mechanisms* by demonstrating how emerging metrics offer complementaries to the existing ones and are inspired by and shed light on the mechanisms of science. Based on fundamental properties of science, we classify existing theories and findings into: *hot and cold* science referring to attention to scientific fields, *fast and slow* science reflecting productivity of scientists and teams, *soft and hard* science revealing reproducibility of research papers. In extracting these properties, we follow the practices of Derek J. de Solla Price, who viewed science as a complex system and developed the science of science via an analogy to thermodynamics. Specifically, he conceptualized science as a gas with individual molecules (scientists) possessing velocities and interactions and exhibiting general properties or laws (Price 1963). Bringing back the emphasis on the properties and dynamics of science shared by Price, Robert K. Merton, Eugene Garfield, and many others may complement the zeitgeist in pursuing new, complex metrics without understanding the underlying processes.

We are not suggesting the abandonment of metrics. Instead, we join a growing number of scholars who argue for the careful study and cautious use of metrics (Hicks et al. 2015). This is critical because each metric comes with a hidden "worldview" of science, and how we measure science affects how science works—as Goodhart's law states, "when a measure becomes a target, it ceases to be a good measure" (Goodhart 1984). In particular, by linking metrics to mechanisms, we hope to motivate broad research interest in analyzing dynamic processes by scholars in the Science of Science, Scientometrics and Informetrics, Science and Technology Studies, and beyond. Our objective is to motivate a more conscientious use of metrics by research institutions, funding agencies, and policymakers based on understanding what society can expect and get from science (Bernal 1939). In this way, we propose contextualized and theory-derived quantitative metrics designed to expose the currently unmeasurable structures and mechanisms of science.

**Why now?**

Metrics and mechanisms of science were connected from the beginning. When Derek J. de Solla Price envisioned and founded scientometrics, he called it "Science of Science" to emphasize the vision of this new field as turning scientific curiosity and research methods onto ourselves to understand the properties, laws, and mechanisms of science (Price 1963). Price wasn't doing this alone but worked together with allies from other disciplines, including Robert K. Merton, who founded the sociology of science (Merton 1973), and Eugene Garfield, who created the Science Citation Index, the current Web of Science (Garfield 1964). Over time, significant focus on evaluative metrics, especially ones based on citation, have left some of the early questions on social mechanisms understudied. Since then, with increasingly available data and computation, the time has ripened for this vision to be fully realized.

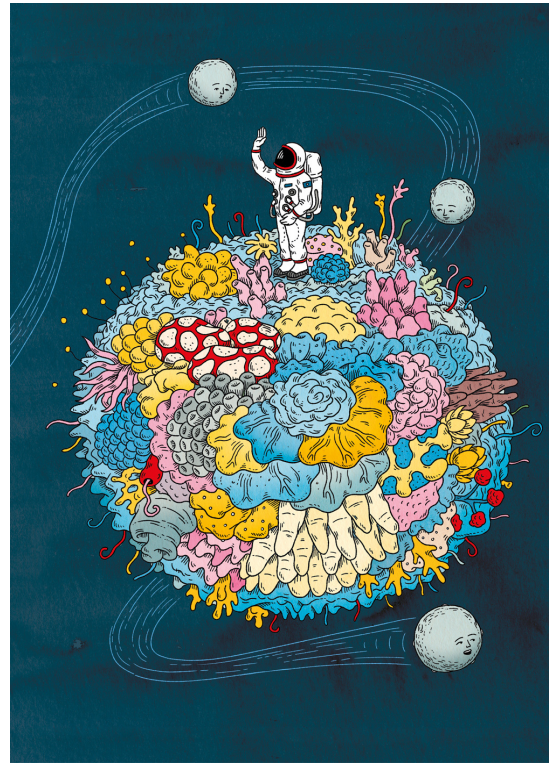

Figure 1. In an age of big data and smart machines, the Science of Science research once dreamed by Price and others has now become a reality—we can explore the "deep space" of knowledge initiatives and measure the previously unmeasurable structure and dynamics of science.
ILLUSTRATION: ANDREW RAE

As scientific enterprises continuously drive the 21-century knowledge economy, it has become apparent that the ambitious inquiries raised by Price and his extraordinary colleagues are unlikely solvable within any discipline alone. The recent Science of Science revival (Fortunato et al. 2018; Wang and Barabási 2021) has provided a new array of whys, wherefores, and know-how to expand the scope and depth of scientometrics by integrating theories and methods from sociology, economics, management science, psychology and beyond. These include the social process of knowledge production (Merton 1968; Foster, Rzhetsky, and Evans 2015), the economics of scientific idea production (Ahmadpoor and Jones 2017; Azoulay, Fons-Rosen, and Zivin 2019), the science of organization and teams in collective innovation (Fiore 2008; Woolley et al. 2010; Boudreau, Lacetera, and Lakhani 2011; Lorenz et al. 2011; Mason and Watts 2012; Cummings and Kiesler 2014), and the psychological structure of the talent and creativity of individual scientists (Simonton 1979; Liu et al. 2018), just to name a few.

These interdisciplinary initiatives have greatly expanded the creative potential of possible future research in an age of big data and smart machines. The research once dreamed by Price and others has now become a reality (Figure 1). The new landmarks include complex network models applied to uncover "invisible colleges" (Crane 1972; Newman 2001), machine learning models developed to retrieve "undiscovered public knowledge" (Swanson 1986; Tshitoyan et al. 2019; Kittur et al. 2019; Peng et al. 2020), novel datasets used to test the proposition that "science progresses one funeral at a time" (Planck 1950; Azoulay, Fons-Rosen, and Zivin 2019). The expanding list also includes online repositories of discoveries investigated to quantify "multiple discoveries" (Merton 1961; Hill and Stein 2019; Painter et al. 2020), citation graphs analyzed to measure "paradigm shift" (Kuhn 1962; Chen 2004; Wu, Wang, and Evans 2019), and computational linguistic models employed to illuminate the formation of scientific consensus (Bruno Latour 1987; Shwed and Bearman 2010; Jurgens et al. 2018). These are merely a representative sample of studies that we use to argue for a more systematic and broad consideration of metrics and how they relate to the hidden mechanisms of science. In the following sections, we discuss how a meta-level perspective allows us to demonstrate the complementarity of these metrics to produce a more integrative and, sometimes, holistic understanding of the scientific ecosystem.

## 2. The Properties of Science

In 1963, Price made the beautiful analogy between the science of science and thermodynamics, conceptualizing science like gas with individual molecules (scientists) possessing velocities and interactions, a total volume, and general properties or laws (Price 1963). We extend Price's analogy to find out more properties that can be analyzed using the paradigms of physics, such as *temperature*, *velocity, and density*. Based on these properties, we classify existing theories and findings into: *hot and cold* science referring to attention to scientific fields, *fast and slow* science reflecting productivity of scientists and teams, *soft and hard* science revealing reproducibility of scientific research. We show how emerging metrics complement existing ones and reveal underlying social mechanisms for a representative list of research topics in the Science of Science (Table 1).

Table 1. The emerging, complementary metrics as the products of theories on the mechanisms of science.

| 2. The Properties of Science | | | |
|---|---|---|---|
| **Topics** | **Metrics** | | **Mechanisms** |
| | Existing | Emerging | |

| 2.1 Hot & Cold Science | | | |
|---|---|---|---|
| *2.1.1 Discovery Significance* | Citation number | Disruption Score (Funk and Owen-Smith 2017; Wu, Wang, and Evans 2019) | *Theory Creation and Destruction*<br><br>Scientific revolution (Kuhn 1962)<br>The creative destruction of science (Schumpeter 1942; McMahan and McFarland 2021)<br>The burden of knowledge (Jones 2009) |
| *2.1.2 Concept Establishment* | Keyword popularity | Meme score (Kuhn et al., 2014)<br>Ambiguity score (McMahan and Evans 2018)<br>Citation function (Jurgens et al. 2018)<br>ForeCite score (King et al., 2020) | *Concept Explicitness and Ambiguity*<br><br>The black box of science (Bruno Latour 1987)<br>Scholarly consensus (Shwed and Bearman 2010) and debate (McMahan and Evans 2018) |
| *2.1.3 Discipline Formation* | Keyword age | Intellectual turning points (Chen 2004)<br>Paradigmaticness (Evans et al., 2016)<br>Rao-Stirling index for interdisciplinary (Leahey et al., 2017) | *Discipline Independence and Integration*<br><br>The republic of science (Polanyi 1962)<br>Rapid discoveries (Collins 1994)<br>Discipline cohesion (Moody 2004)<br>Boundary-spanning (Sebastian and Chen 2021) |
| *2.1.4 Research Impact* | Citation number | Eponymy (McCain 2011)<br>Forgotten Index (McGail 2021)<br>Sleeping Beauty Index (Ke et al. 2015) | *Knowledge Forgetting and Remembering*<br><br>Obliteration by Incorporation (Merton 1968; McCain 2014)<br>Delayed Recognition (Garfield 1980)<br>Sleeping Beauty (van Raan 2004) |
| **2.2 Fast & Slow Science** | | | |
| *2.2.1 Scholar Performance* | H-index (Hirsch 2005) | Q parameter (Sinatra et al. 2016) Credit allocation algorithm (Shen and Barabási 2014) | *Scholarly Creativity and Recognition* <br><br> The Matthew's effect (R. K. Merton 1968) <br> Invisible college (Price and Beaver 1966; Crane 1972) |
| *2.2.2 Knowledge Cumulativeness* | Number of papers | Atypicality (Uzzi et al. 2013) <br> Pivot size (Hill et al. 2021) <br> Cognitive extent (Milojević 2015) <br> Price Index (Price 1969; Milojević 2012) | *Adjacent Possible and Realization* <br><br> Adjacent possible (Kauffman 2000; Johnson 2011; Youn et al. 2015; Loreto et al. 2016) <br> Rapid discoveries (Price 1969; Collins 1994) |
| *2.2.3 Concept Relevance* | Bibliographic coupling | Knowledge graph (Bordes et al. 2013) <br> Analogy modeling (Tshitoyan et al. 2019; Kittur et al. 2019) | *Shallow and Deep Association between Knowledge* <br><br> Undiscovered public knowledge (Swanson 1986) <br> Analogy as the fuel and fire of thinking (Hofstadter and Sander 2013) |
| **2.3 Soft & Hard Science** | | | |
| *2.3.1 Finding Replicability* | Number of figures | Embedding dimensions of soft vs hard science (Peng et al. 2021) | *Soft and Hard Science* <br><br> Hierarchy of the sciences (Price 1969; Cole 1983) <br> Drawing things together (B. Latour 2011) |
| *2.3.2 Discovery Originality* | Anecdotes | Priority race identification (Hill and Stein 2019; Painter et al. 2020) | *Singletons and Multiples in Scientific Discoveries* <br><br> Multiple discoveries (Merton 1961; Simonton 1978; Brannigan and Wanner 1983) |
| *2.3.3 Knowledge Spreadness* | Citation number | Product space (Hidalgo et al. 2007) <br> Effective travel distance (Brockmann and Helbing 2013; Coscia et al., 2020; Catalini et al., 2020) | *Tacit and Codified Knowledge* <br><br> Tacit knowledge (Polanyi 1966) <br> Sticky, locus information (von Hippel 1994) |

## 2.1 Hot and Cold Science

Heat is a property of molecular kinetic energy, and, analogously, emerging fields, where science is still "in the making" through debates, are hotter and of higher vitality than established fields. There is less activity, or "cold," where scholarly consensus has formed, and science is "made." Latour referred to the "black box" of scientific concepts to capture the idea of codified knowledge that has been created, sealed, and passed down (Latour 1987). The "black box" may be opened again in the scientific revolution (Kuhn 1962), when the cold, normal science becomes hot and vital again.

Here, we can continue with our physics metaphor with the notion of heat as thermal energy transfer when systems come in contact. Energy is transferred from a hotter system to a cooler system when they collide, and this can occur in science for many reasons. In some instances, scholars from different disciplines actively engaged in new theory or methods (hot science) can explore their ideas in older or established areas (cold science). Or, novel paradigms can be created by "newcomers," the new generation of scholars or scholars from other fields, who are more innovative because they are less inclined to accept the 'black box' of passed down knowledge, or simply know less and, therefore, do not have the "burden of knowledge" (Jones 2009). Of course, in many cases, "premature" ideas may appear ahead of their time and are therefore forgotten. Some of them are lucky to be remembered again and enjoy delayed recognition (Garfield 1980), and that contact may then transform a cold science into a hot science. Through the dynamics of hot and cold seasons of science, paradigms are created and destroyed, claims are established and challenged, disciplines celebrating independence are melded with other disciplines, and scholars creating similar work are amalgamated. Next, we discuss the quantities of interest we see as important for understanding the dynamics of and between hot and cold science. We do this through analyses of existing and emerging metrics and describe the mechanisms through which these can alter science.

### 2.1.1 Discovery Significance: Creation and Destruction

It is a human tendency to separate decisive moments from trivial ones and remember major historical events instead of daily routines. We ask, "what are the most significant scientific discoveries?" But we often forget that science is perpetually in flux. Science is the ever-changing river of Heraclitus, where one can not step twice into the same circumstance. The *creative destruction of science* (McMahan and McFarland 2021), in which new ideas substitute the old, has been discussed under different names but shared sentiments (Polanyi 1962; Bourdieu 1975; March 1991; Whitley 2000), among which the "*normal science*" and "*revolutionary science*" proposed by Kuhn are perhaps the most widely used (Kuhn 1962).

The fact that ideas are created and destroyed repeatedly reminds us that research consolidating existing scientific paradigms needs to be examined in the context of work that challenges those paradigms with a purpose to create new paths. The latter is typically led by "newcomers," the new generation of scholars or scholars from other fields. These scholars are less familiar with existing theories and thus have no *burden of knowledge* (Jones 2009), or they may be aware but disagree with the established way of thinking (Kuhn, 1962). To illustrate how these metrics can similarly guide science, they could be incorporated more systematically into graduate education. For example, scientists typically only receive exposure to perspectives on knowledge creation or discussion if they take a course on the history and philosophy of science. By clearly linking these to quantifiable metrics, these ideas could be incorporated into traditional discipline-based courses to illustrate the progression of knowledge. This may motivate each generation of scholars to be less accepting of prevailing perspectives and to open the "black box" of science.

Recent research distinguished between two different scientific activities, "developing vs. disruption," using a "*disruption*" measure or *D*-score (Funk and Owen-Smith 2017; Wu, Wang, and Evans 2019). The intuition behind D-score is straightforward: if subsequent papers that cite a paper also cite its references, then the focal paper can be seen as consolidating the prior knowledge upon which it builds. If the converse is true—future papers citing a focal paper ignore its acknowledged forebears—they recognize that paper as disruptive, creating an unanticipated new direction for science. Specifically, D-score is calculated as the difference between the fraction of these types of subsequent, citing papers. Two works may have the same citation counts but very different *D*-scores. For example, the BTW-model paper (Bak et al., 1987) that discovered the "self-organized criticality" property of dynamical systems, one of the most prominent patterns in complexity science, is among the most disruptive papers ($D = 0.86$, top 1%). In contrast, the article by Davis et al. (1996) on first observing the Bose-Einstein condensation in the lab is among the most developing ($D = -0.58$, bottom 3%). Our point is that both were published in the same prominent journal, *Physical Review Letters*, and both have over 8,000 citations, according to Google Scholar. But it is their *D*-scores that highlight the distinct nature of the knowledge created by these two papers. The *D*-score is just an example of many possible measures that can be defined from the citation network to quantify scientific advances in steps or leaps, and it has its limitations.

*D*-score compresses two dimensions, i.e., "developing" and "disrupting," into one variable to maximize the discriminant power of that single variable. Recent work proposed to unfold these two dimensions to capture the "dual characteristics" of technology (Chen et al., 2021). Emerging studies also engaged in evaluating and developing *D*-score by linking it to paper content novelty measured through keywords (Leahey et al. 2021) or references (Lin, Evans, and Wu 2021) or measures of "interdisciplinarity" and "synergy" in scientific collaborations (Leydesdorff and Ivanova 2021). Future research can also systematically test the relationship between disruption and novelty (e.g., Uzzi et al., 2013) to understand whether they capture different kinds of scientific advances. Indeed, the D-score can also help the process of science itself by aiding one's literature review. For example, D-scores can guide a scientist's reading of the literature to seek breakthroughs on fundamental problems within the field and learn from the best practices.

### 2.1.2 Claim Establishment: Consensus and Debate

The "*black box*" (Bruno Latour 1987) metaphor helps us better understand the turn-taking between "normal science" and "revolutionary science" from the perspective of scientific products. Scholars conducting normal science take existing claims for granted to develop further assumptions, and scientific revolution happens when the new generation of scholars try to open the "black box" of sealed claims and challenge it with new ideas or conflicting evidence. In this way, the creation and destruction of scientific claims are achieved through scientific workforce substitution (Azoulay et al. 2019).

This does not necessarily deconstruct scientific facts but reminds the importance of the social processes through which we collect these facts. Indeed, a scientific claim such as "smoking causes cancer" is built on *scholarly consensus* formed through a long, complicated process before they are widely accepted (Shwed and Bearman 2010). Most scientific concepts can be traced to a single article that dominates the citations and is recognized as the origin. The *ForeCite* algorithm (King et al. 2020) developed from this observation is highly effective in distinguishing scientific concepts (e.g., "fast gradient sign method") from common words (e.g., "effective method").

To better position the evaluation of metrics on claim establishment, we need to look at the temporality of change. Each scholarly consensus is typically preceded by substantive debate about a concept or established finding. We suggest that research needs to more carefully attend to the nature of the concept and the context of its use. For example, concept *ambiguity* (Chen et al. 2018) may lead to a higher level of engagement than concept clarity (McMahan and Evans 2018). In the latter, there is a convergence of meaning that facilitates the propagation of ideas as *scientific memes* (Kuhn et al. 2014) and may make it less attractive to innovators. More importantly, to consider contextualized, detailed study of scientific debates, machine learning models can help analyze *citation functions* (Catalini et al. 2015). What is critical for understanding citations in context is that, contrary to the common belief that people mostly cite to agree, nearly 17% of citations are for "comparison or contrast." Yet, these citations attract less attention (Jurgens et al. 2018; Lamers et al. 2021). Here we have noted another example of how to analyze the literature space more carefully while also guiding the creation of this space. For example, scholars can include citations that conflict more regularly, so the breadth of some literature space is continually conveyed. Although this might forestall consensus, it may increase the robustness of any concept that does get established (Chen 2020).

### 2.1.3 Discipline Formation: Independence and Integration

Polanyi (1962) created the concept of the "*republic of science*" to argue that the practicing community of science implicitly organized itself in ways similar to political and economic systems, which is contradictory to Bernal's argument that scholars and policymakers can and should plan science explicitly (Bernal 1939). By making Polanyi's point clear, scientists can better manage their actions and interactions in more planful and productive ways. For example, the republic of science expands its territory and grows new disciplines through interactions between the new and the old, including methodological fields that export broadly. Emerging topical fields borrow heavily to expand, whereas old topical fields grow insular and retract (Ramage et al. 2020).

In this process, science presents two faces: *science in the making* and *science-made* (Latour 1987). Science in the making is chaotic, dynamic, with heated, "fluid" knowledge spilling over from one discipline to another, whereas science made is relatively more organized and stabilized, with cooled, "crystalized" pieces of claims hanging in the hall of frames, worshiped by "disciplined" scholars (Sugimoto and Weingart 2015).

So what are the key factors that lead to the creation or growth of disciplines? Investigating 16th-17th century Europe under the Scientific Revolution may help address this question. At this period, several branches of natural science shifted into a fast track of independent development characterized by *rapid discovery and high consensus* (Collins 1994). For example, chemistry became independent from alchemy after the work of Robert Boyle, and astronomy gained autonomy from astrology after the work of Galileo Galilei. In contrast, metaphysics, promoted by Francis Bacon, or social physics, advocated by Thomas Hobbes, failed to become independent from philosophy.

Why was this so? On the one hand, it could have been because of Bacon and Hobbes' lack of talent, passion, or connection compared to Boyle and Galileo. On the other hand, external factors could have driven the developments. Setting aside differences due to training of scientific methods, or the complexity of research topics, Collins (1994) proposed that research technology was the key to making science "take off." One could argue that we can better understand the development of scholarly consensus by analyzing the genealogy of research technologies (e.g., telescopes in 16th-17th century astronomy). These research tools lower the bar of training new scientists, inexpensively produce new facts, and help scientists within the field rapidly form a consensus on old topics to adapt to the moving research front. By developing metrics to track such changes, we may be better positioned to identify where science funding needs to be focused (e.g., on infrastructure) to accelerate developments.

In the "science in the making" stage, there could be *competing paradigms*, trackable through the density curves of citations (Chen 2013). After science is made and *discipline cohesion* is achieved (Moody 2004), high consensus measured in entropy of keywords, and rapid discovery measured in the number of emerging papers, can be used to quantify the formation and stabilization of paradigms or *paradigmaticness* of a discipline (Evans et al., 2016).

As the gap between disciplines increases, it takes a Renaissance person or an interdisciplinary team to bridge it. However, the cognitive cost to go across the boundaries is high. *Individual concept maps* are typically very different and only overlap marginally between scholars trained from different traditions (Gowin and Novak 1984; Novak and Cañas 2006; Dai and Boos 2019). Therefore, the collaboration between scientists across disciplines is challenging, as they face a challenge of concept integration to bridge these gaps to produce new knowledge (Salazar et al. 2012). The *Rao-Stirling index* (Leahey, Beckman, and Stanko 2017) is a good way to measure the cost as it considers the spanning of disciplines and the evenness and cognitive distance between these disciplines.

Effective discipline integration comes with a price, making the author(s) more prominent in the long term but less productive in the short term (Leahey et al., 2017). For this reason, nominal interdisciplinary collaborations in which scholars strategically team up for grant opportunities (Dai 2020) or shallow interdisciplinary collaboration driven by labor division (Haeussler and Sauermann 2020) of hierarchical teams may happen. To distinguish "flat" teams in which multiple members play leadership roles from "tall" teams dominated by a small number of leaders, Xu et al. (2022) proposed "lead (or $L$)-ratio" as the fraction of team members engaged in conceptual work (e.g., design research, in compared to technical work such as analyzing data) by analyzing author contribution reports. They found that tall, hierarchical teams produce less novelty relative to flat, egalitarian teams and more often develop existing ideas; increase productivity for those on top and decrease it for those beneath; increase short-term citations but decrease long-term influence (Xu et al. 2022). This metric, then, can also identify areas of research that need deeper expertise and help in graduate education and professional development, target scholars and policymakers in these areas to ensure that genuine interdisciplinarity can emerge and be supported.

### 2.1.4 Research Impact: Forgetting and Remembering

As already discussed, the science citation index allows us to track the genealogy of ideas (Schoenbach and Garfield 1956; Garfield 1964). However, citations only, and not surprisingly, tell us what has been cited. Related citations may have been left out due to space constraints, author ignorance, or, although relevant, no longer seen as needed. About the latter, one mechanism challenging the conventional analysis of citations is "*obliteration by incorporation*" (OBI): when ideas, methods, or findings are incorporated into currently accepted knowledge and treated as "common sense," their source is obliterated (Merton 1968). No contemporary physicists would cite Newton's Principia to refer to the law of universal gravitation. As a result, as more papers use original work, the number of explicit references ironically declines. In the successive transmission of ideas across generations, different versions of science are integrated into the legitimated, centralized one, creating a palimpsestic replacement (McCain 2014). All but the immediately antecedent versions are faded, and citations hardly precisely point to the ideas' true origins. Furthermore, this mechanism is moderated by information and communication technologies such as Google Scholar. Searching online is extremely efficient, and following hyperlinks quickly puts researchers in touch with prevailing opinions, but this may accelerate consensus and narrow the range of findings and ideas built upon (Evans 2008).

In many cases, papers are forgotten but then remembered again. This phenomenon was called "premature" discoveries in the 1960s (Barber 1963; Stent 1972; Wyatt 1975; Hook 2002), "delayed recognition" after the 1970s (Cole 1970; Garfield 1989, 1990), and "revived classics" or "sleeping beauties" after 2000 (Krapivsky and Redner 2001; van Raan 2004; Ke et al. 2015). For the origins of delayed recognition, Barber (1963) suggested that resistance to new discoveries partially stems from the "tendency of scientists to think in terms of established models" and "repulsion to potentially change." This bias against novelty happens across disciplines (Wang, Veugelers, and Stephan 2016). For the frequency of delayed recognition, Glänzel and Garfield observed that most papers are well cited within the first three to five years of publication, except for a small fraction of outliers—0.01% according to their study—experiencing a burst of attention after ten years (Glanzel and Garfield 2004).

Recent studies with large-scale data suggest that these "outliers" may be more common than expected (van Raan 2004; Yin and Wang 2017; He et al., 2018). Based on the analysis of the "*sleeping beauty index* (SBI)," a non-parametric measure, papers among the top 0.1% SBI still demonstrate a clear pattern of delayed recognition, ten times larger than what Glänzel and Garfield found (Ke et al. 2015). One possibility is that Glänzel and Garfield analyzed papers published only before 1980 and so missed the opportunity to discover many papers "awakened" in recent decades. Nonetheless, this finding identifies a critical issue that Glänzel and Garfield ignored—papers of belated recognition may be rare, but this does not mean that they are unimportant. On the contrary, these papers are likely too novel to be recognized immediately, and their importance unfolds over time. As for the function, half of "sleeping beauties" are research technologies. Therefore, they are "sleeping innovations" (van Raan 2017). Furthermore, the use of these technologies outside science, measured in patent citations, may come after their "awakening" within science (van Raan 2017). An interesting question here, though, is whether papers of delayed recognition will be more likely to appear over time. On the one hand, scientists are searching deeper into the past measured in reference age (Cui et al., 2022), which decreases the chance that valuable literature is ignored; on the other hand, scientists are increasingly specialized (Jones 2008). Therefore, relevant discoveries from other fields may sleep long before being awakened by a polymath. This paradox points to a future direction of delayed recognition studies.

There have been several attempts to measure the OBI mechanism, including the cases of Nash Equilibrium (McCain 2011) as summarized together with other attempts (McCain 2014). For now, the most established measure is *eponymy:* comparing the degree to which implicit citations are or are not included when including the named concept without the author (McCain, 2011). The recently developed *Forgotten Index* can also be used to infer the existence of the OBI effect indirectly (McGail 2021). Finally, because such metrics can help us understand how citations diminish or disappear, they may also help identify the drivers of innovation. In particular, new findings in a given area may benefit from seeking out lost citations as these earlier papers may have additional data or concepts to increase the impact of the newly discovered knowledge - essentially seeing something old with a fresh perspective.

## 2.2 Fast & Slow Science

Different branches of science move at different paces. Research technology plays a key role in forming sciences characterized by high consensus and rapid discoveries (Collins 1994, 2009). When science moves forward, it incorporates existing knowledge so that the common knowledge is used without citations (Merton 1968; McCain 2014). Therefore, the "Price index," the fraction of recent references against all references, measures the pace at which science moves.

In theory, science should move at an accelerating pace, as the emerging ideas, concepts, and methods will exponentially expand the space of "adjacent possible," including all possible ideas as the combination of existing elements (Kauffman 2000; Johnson 2011; Youn et al. 2015; Loreto et al. 2016). But analyzing the growth of total knowledge must account that many individual scientists and teams are working on evaluating rather than searching for new ideas. Thus the total number of papers is a misleading measure that favors productivity at the surface, compared to the total number of unique concepts that reflect the "cognitive extent" of science (Milojević 2015). Will science hit its upper bound, as Price conceived in 1963 and recently emphasized by the "ideas getting harder to find" (Bloom et al. 2020)? Or, will science move towards the "endless frontier" (Bush 1945) with an increasing rate of realizing possible ideas, as promised by the recombinant theory (Arthur 2009)? And how would AI play a role in this process? These are pressing questions in our time that are calling for answers. We next describe related quantities of interest and their associated metrics that may help address this question.

### 2.2.1 Scholar Performance: Creativity and Recognition

The performance of scholars is perhaps the quantity that generated the most interest among scientists and the broadest range of indicators. The complexity of this issue originates from the decoupling between the quality and recognition of artifacts due to social influence (Salganik et al. 2006). In particular, the "*Matthew effect*" in science, in which the initial, random differences between scientists in status amplify over time, results in the increasing cumulative advantage of successful scholars (Price 1965; Merton 1968; Wang et al. 2013). The positive feedback mechanism transforms even a normal distribution of inherent aptitude among scientists into the long-tail distribution of recognition.

While citation impact and its derivatives such as *H-index* (Hirsch 2005) are biased products of social influence and do not reflect a scientist's "intrinsic" creativity, the *Q parameter* allows inferring creativity after controlling for productivity and social influence (Sinatra et al. 2016). This control is important as normalization of citations across fields is crucial (Ioannidis et al.

2019), and the journal impact factor is also sensitive to journal size and outlier effects due to the Central Limit Theorem (Antonoyiannakis 2018, 2020). Other metrics probe deeper to consider the role of individuals in multi-authored papers. For example, the *credit allocation algorithm* (Shen and Barabási 2014) tracks the recognition of author contributions within each paper over time. This algorithm accounts for scholar visibility within the scientific community. Specifically, one needs to stay in the discipline long enough to collect credit, as an author's credit in a specific paper is not a fixed value but dynamically evolves, increasing with the author's subsequent publications. This process creates another layer of the Mathew effect that favors scholars who survive longer (Petersen et al. 2011) and against temporary scientific workers (Milojević 2018).

The decoupling between creativity and recognition reminds us that social context is critical. Although these emerging measures can add new dimensions to the measurement of the individual or team performance, we should be keenly aware that metrics themselves are not certainly free from systematic biases against underrepresented populations in science and, therefore, cannot automatically address deeply rooted systematic inequality.

### 2.2.2 Knowledge Cumulativeness: Adjacent Possible and Realization

Science grows by accumulating knowledge based on scholarly consensus and moving forward to include new ideas and evidence. This observation brings the question of how fast a research front moves and what factors determine it? There are two theories on how science advances. The more optimistic one is derived from the exponential growth of scientific literature, which doubles every 15-20 years (Price 1963). As described earlier, an explanation for this growth is the theory of *adjacent possible.* This theory assumes that new ideas are the combinations of existing ideas. Therefore, the space of all possible new ideas as the combination of existing elements should increase exponentially with the pool of available elements (Kauffman 2000; Johnson 2011). This prediction is particularly aligned with periods of *rapid discoveries* (Collins 1994) in science, including the 16th-17th century of Europe when the scientific revolution was happening, and perhaps also after World War II in the U.S., when it was believed that science would move toward the "endless frontier" (Bush 1945).

In contrast to the more optimistic view, the *upper bound of science* predicts that science may hit a wall, and the exponential growth will collapse into a logistics curve (Price 1963). Recent studies echoed this concern on stagnant science (Collison and Nielsen 2018) and the diminishing returns of science (Bloom et al. 2020). Measuring the pace at which the research front moves by the *Price Index*, which is the fraction of new references (Cui et al., 2022) and disruption score (Park et al. 2021), also suggests that science is slowing down. This slow increase of science is also supported by the linear growth of new concepts (Milojević 2015) and technology codes (Youn at al. 2015) in contrast to the exponential growth of papers and patents. The direct comparison between adjacent possible and realization in patent inventions (Youn et al. 2015) and scientific discoveries (Loreto et al. 2016) confirmed that only a small fraction of all possible ideas are created—science seems to move slowly and conservatively. Cui et al. (2022) associated the slowed advance with the aging scientific workforce. But more work can be done to unfold the underlying causes. A critical issue with the use of these measures is to identify how the rate of growth relates to science funding. If declines in innovation and increases in funding coincide, reflection upon funding strategies is necessary for better science planning. Improved metrics linked to more substantive innovative processes can more productively inform ways to support science.

### 2.2.3 Concept Relevance: Shallow and Deep Association

The recombinant theory predicts an accelerating growth of science, as each new idea expands the creative potential of all possible ideas (Arthur 2011). On the contrary, the social network theory of science argues for a slow expansion of science, as science is embodied in scientists and new ideas need to be "introduced" by old ideas (Collins 2009). To this end, an effective strategy of proposing something new is to make it a "familiar surprise" by anchoring it to the well-accepted ideas (Kim et al. 2016). The best strategy to write a hit paper is to start from highly traditional knowledge and try to add novel ingredients, as demonstrated by Uzzi et al. (2013) using a score of *typicality* (Z-score), which identifies whether co-cited journals appear more or less likely than random in the literature. Moving from this less risky strategy of innovation towards the more risky end involves creative leaps to link distant ideas, such as how Lilian Bruch connected "HIV" with "monkeys" to introduce HIV's origins in nonhuman primates (Sarngadharan et al. 1984), or how Londa Schiebinger (1991) linked "masculinity" to "justify" and pioneered the academic studies on gender bias. These innovations are long shots for career success as it trades immediate productivity for long-term impact (Leahey et al. 2017), and are disproportionately contributed by minority groups (Hofstra et al. 2020), small teams (Wu et al., 2019), and young scientists (Packalen and Bhattacharya 2015). It is important to understand better why these demographics are more likely to be innovative to improve the likelihood of accelerating innovation and offset the slow pace of science.

The recombinant theory also assumes that existing knowledge and technical components are equally likely to complement each other and form new ideas. But what if existing components are highly redundant? There is evidence that 2,000 discourse atoms in English are enough to represent over 100,000 words that capture different meanings (Arora et al. 2018). To this end, quantifying the deep association between scientific concepts beyond co-occurrence is the premise of understanding the potential of science.

Swanson had envisioned that science could advance by linking knowns to reveal unknowns: "... independently created pieces of knowledge can harbor an unseen, unknown, and unintended pattern. And so it is that the world of recorded knowledge can yield genuinely new discoveries" (Swanson 1986). But his best practice still relied on keyword co-occurrence to discover hidden associations. An example was that Raynaud's disease and fish oil were both related to blood viscosity, which predicts that dietary fish oil could be used to cure patients with Raynaud's disease, as confirmed by subsequent clinical trials (Swanson 1986).

A substantial advance from keyword co-occurrence analysis is to model analogy, which answers questions such as "a is to b as c is to _?" Artificial intelligence theorists have argued *analogy as the fuel and fire of thinking* (Hofstadter and Sander 2013). Cognitive science has studied how the analogy is used in research laboratories and found that labs finding disparate connections are more likely to do so via analogic reasoning (Dunbar 1997), and that within-domain, versus cross-domain analogies, will differentially drive innovation (Paletz, Schunn, and Kim 2013). Unlike these implicit analyses of analogy, the emergence of *knowledge graphs* and Translational embedding (Bordes et al. 2013) explicitly model the "a is to b as c is to _?" question and embed entities and relations into low-dimensional spaces for scientific reasoning tasks. Examples include predicting new thermoelectric materials (Tshitoyan et al. 2019) or early singles for impactful research (Weis and Jacobson 2021). Others use machine learning to determine the kinds of questions that need to be asked and answered based upon suggestions, gaps, and speculation (Malhotra et al. 2013). Machine learning models also effectively extract and represent "problems" and "solutions" in research articles and facilitate productive analogies for discovering new solutions to existing problems (Chan et al. 2018). These metrics, and the subsequent tools based upon them, represent important developments for helping automate or accelerate scientific innovation (Kittur et al. 2019). It may be that developments in AI can overcome limitations in slow science by accelerating either discovery or consensus (Collins 1994). This process is comparable to how network science and computer science, combined with sociology, created computational social science to accelerate research in the detailed study of human behavior (Lazer et al. 2009). Regardless, metrics-based upon novel interpretations of existing knowledge, whether through analogy or integration, can help science policy and practice better understand how to address problematic areas where science is slowing.

**2.3 Soft and Hard Science**
Largely inspired by the *unequal evolution of scientific knowledge* (Nelson 2005), scholars have sought to identify the cognitive status of the sciences and differentiate between "hard" and "soft" science. Philosophers of science attempt to ground such distinctions in abstract principles, and sociologists of science seek to identify relevant differences (for example, in the level of social consensus). Recent investigations of scientists' concrete data representation practices provide new leads on this problem. Bruno Latour has argued for the essential role of graphs in producing scientific facts due to their ability to "draw things together," rendering phenomena into a compact, transportable and persuasive form (Latour 2011). Following Latour's intuition, Smith et al. (2000) associated the use of graphs with hardness across seven scientific disciplines and ten specialty fields in psychology. Papers in hard science tend to use more graphs than soft science. Cleveland (1984) surveyed 57 journals and revealed that natural science journals use more graphs than mathematical or social science journals. An interesting question here is whether graphs as a proxy of science hardness are associated with "multiple discoveries," in which two or more scientists independently claim the same discovery. The hypothesis is that if graphs, together with other scientific artifacts (e.g., equations), can serve as devices in rendering phenomena into "densified" knowledge, they may increase the chance scientists converge thinking and "step on each other's toes" for the next discovery, leading to more multiple discoveries in hard sciences than in soft sciences. Following the same intuition on the knowledge codifying function of graphs and equations, we may ask whether hard sciences are more reproducible than soft sciences. We next unfold the discussions on soft and hard sciences in detail by introducing the links between metrics and mechanisms.

### 2.3.1 Finding Replicability: Soft and Hard Science

The pursuit and evaluation of replication and reproducibility have become part of the most important agendas of science (Munafò et al. 2017). Meanwhile, the unequal evolution of meta-analysis across scientific fields reminds us that meaningful discussions on this topic should be conditional on awareness of the distinct nature of knowledge across various domains. Two different perspectives have been developed around this issue. On the one hand, Auguste Comte and many others have developed theories on *the hierarchy of the sciences* to differentiate sciences at the top, which presumably manifest higher levels of consensus and more rapid rates of advances than sciences at the bottom. On the other hand, sociologists of science suggested a distinction between the "core" and the "research frontier" and argued that no systematic differences exist between sciences at the research front (Cole 1983). This echoes the "more is different" theory proposed by leading scientists to reflect upon the hierarchy of science (Anderson 1972).

With arguments from both sides presented, understanding the unequal evolution of scientific knowledge is an intriguing topic [(Nelson 2005)](). For scholars seeking to quantify the "hardness" of sciences, the *number of figures* could be a good starting point to represent "hardness" as introduced above (Cleveland 1984; Smith et al. 2000). Meanwhile, it is important to identify metrics that reflect the semantic content but are agnostic to discipline, therefore, are robust against biases introduced by writing practices. The emergence of large scientific corpus and machine learning methods allow us to construct knowledge spaces from the association of journals and discover the *hidden dimensions of soft vs. hard science* (Peng et al. 2021) at continuous models instead of graphs (Boyack, Klavans, and Börner 2005). Building upon these metrics, it might be feasible to target science policy in areas of science needing more attention on knowledge integration, that is, drawing concepts together and on methods to improve replication.

### 2.3.2 Discovery Originality: Singletons and Multiples

Another important question entangling with the unequal evolution of sciences is whether science is "inevitable"? Or, putting it more specifically focusing on individual performance, whether creativity in science originated from *genius or zeitgeist* (Simonton 2004)? Merton first proposed this question and hypothesized two possible paths of science, either progressing through "*multiples*" or "*singletons*" (Merton 1961; Simonton 1978; Brannigan and Wanner 1983; Painter et al. 2020). The most famous examples include the debate on the priority over integrals between Newton and Leibniz, or the theory of evolution of species, independently advanced in the 19th century by Charles Darwin and Alfred Russel Wallace. This phenomenon overshadows the pursuit for "originality" of scientific ideas and complicates the precise reference of the origins of scientific ideas.

Due to the challenge of identifying multiple discoveries, previous research relied on anecdotes or hand-crafted lists of multiples to conduct analysis. Despite the limited data, the inevitability of science—at least some parts of it, seems to be supported. Thirty percent of discoveries will see their multiples within a year; 70% of multiples happen within a decade (Merton 1961). According to the Poisson distribution of multiple magnitudes, most discoveries are multiples (Simonton 1978). Over the past four decades, multiple discoveries have happened more often—time interval separating independent reports is diminishing, but less dramatic—involved fewer scientists each time (Brannigan and Wanner 1983). One possible explanation is that this is caused by more efficient coordination within science through increasingly rapid, open publications, making it easier for scientists to learn each other's work and conceive the next step, and more likely to step on each other's toes if they embrace competition. But reassuringly, open access to knowledge also decreases scientists' and teams' chance of investing too much time as the "sunk cost" to an already solved problem. Finally, scientific recognition is not always a "winner takes all" game; the "scooped" research receives less but still substantial (45%) share of total citations, based on a recent work analyzing *priority races* in submissions to Protein Data Bank, a repository of standardized research discoveries in structural biology (Hill and Stein 2019).

### 2.3.3 Knowledge Spreadness: Tacit and Codified

If scholarly consensus is what sciences are built upon and distinct from each other, an important question concerning the social process of consensus, conditional on the presence of the same scientific fact, is how fast this agreement diffuses across time and space. Setting aside that which is yet to be known, one possibility is that common knowledge is only the tip of the iceberg, akin to the stars in a universe full of "dark matter" for which we can not agree or even describe; that is, a vast amount of personal, tacit, and unspoken knowledge of scientists. The unsuccessful attempts of many great scientists to *codify tacit* knowledge reveal the big gap between these two kinds and the fundamental challenge in transferring one another (Polanyi 1966).

The slow diffusion of tacit knowledge, or "know-how," has constrained the economic opportunities of countries as they move very slowly in the *product space* (Hidalgo et al. 2007): one country cannot primarily grow crops one day and launch a moonshot the next. The "*sticky"* tacit knowledge also shapes the development of companies and the spill-over of knowledge (von

Hippel 1994). Despite the importance of know-how as the bottleneck of technology spreading and economic growth, it moves slowly: tools can be shipped, codes and blueprints can be mailed, but know-how is embodied in workers can only be transferred by moving persons: studying abroad, immigration, business travel, etc. In this sense, the *travel distance* (Brockmann and Helbing 2013; Coscia, Neffke, and Hausmann 2020; Catalini, Fons-Rosen, and Gaulé 2020) between research hubs conditions the spreadness of scientific knowledge, in contrast to the common belief that knowledge can be accessed instantly due to the development of information and communication technologies. Identifying metrics that differentiate between tacit and codified knowledge within and across disciplines provides important targets for science policy and education. On the one hand, if disciplines are too low in codified knowledge proportionate to tacit knowledge, science policy can identify research areas for knowledge development. On the other hand, graduate education, within disciplines too high on tacit knowledge, would need to focus research education on ways to improve codification of knowledge.

## 3. Discussions: Science and Society

Science is a complex social activity driven by the efforts of scientists who aim to expand human knowledge by making more things measurable and reasonable. However, with the expansion of science, scientific undertakings at the interface between the natural world and knowledge representations are increasingly complicated, nuanced, and hidden. The manifested features of scientific research are produced by underlying structures and mechanisms. This presents the urgent challenge to develop, test, and implement new science metrics in combination with theories and models that will better capture the properties and dynamics of science. We suggest that understanding the processes of science conditions the effective design, interpretation, and application of the metrics as the outcomes of these processes. Our understanding of the underlying structure (such as how citations distribute), and dynamics (such as how citations grow) of science, scaffolds effective evaluation of outcomes (such as paper citation number, journal impact factor, and scholar h-index) and productive science-policy for societal benefits.

To this end, scientometrics is an underdeveloped, rather than overexploited, field. Perhaps because convenience is where we start, whereas measuring the unmeasurable is a valuable but difficult endeavor. For example, despite numerous metrics on impact within academia we know relatively less about the downstream impact of science in society, such as how basic research supports patent invention (Ahmadpoor and Jones 2017) or policy development (Yin et al. 2021).

In this paper, rather than merely discussing the limitations of widely-used evaluative metrics, we presented a number of possible metrics focused on processes and the resulting structures and built off of theoretical work. We review multiple theories on the mechanisms of science where research creativity is channeled to capture the latent—not immediately explicit—features of science. From this, we argue that using a richer set of metrics inspired by exploration mechanisms may improve the operation of the scientific enterprise as a whole. This can be realized by encouraging different kinds of research—from large to small teams, from developing to disruptive, and from safe to audacious—to be conducted, funded, and rewarded.

At the same time, we acknowledge the drawbacks of quantitative metrics. Such endeavors must always be cognizant that there will always be systemic biases, side effects, or unintended consequences from whatever metrics developed. The wide-spread systematic inequalities and barriers in science disproportionately promote or punish certain intersections of the population, posing a critical challenge to contextualize whatever metrics we employ. Even the most sophisticated metrics with tight connections to the mechanisms cannot easily overcome the fact that the most basic measurements and the system are biased. As numerous scientists have argued (Hicks et al. 2015; Sugimoto 2021), without properly contextualizing the metrics and recognizing the tilted system itself, wielding a metric can do great harm in exacerbating structural inequality in science.

However, this is not an argument for ceasing the development of new metrics. Although metrics provide shortcuts to understanding, an absence of contextualized metrics would lead to heavier reliance on simpler, more problematic metrics. The science of science will, and should, keep moving towards more rigorous yet nuanced measurements and theories that reveal the fundamental dynamics of science while striving to incorporate the rich, complex social contexts that bias everything we measure to *measure the unmeasurable*.

**Acknowledgments**
We acknowledge the following scholars in commenting on this article: Dean Keith Simonton, Julia Lane, Manolis Antonoyiannakis, Ludo Waltman, Cassaundra Amato, Stephen Fitzmier, Richard Van Noorden, Kevin Boyack, Richard Klavans, Jiang Li, Lianghao Dai, Chaomei Chen, Dashun Wang, and James A. Evans. S.M and Y.Y.A acknowledge the support by the Air Force Office of Scientific Research under award number FA9550-19-1-0391. S.M.F acknowledges the support of the National Science Foundation grant 2033970. L. W. acknowledges the support of Pitt Cyber Institute and Richard King Mellon Foundation.